%% file: main.tex
\documentclass[letterpaper, 10 pt, conference]{ieeeconf}  

\IEEEoverridecommandlockouts                              
\usepackage{amsmath}
\usepackage{tikz}
\usetikzlibrary{shapes,arrows}
\usepackage{tikz-network}
\usetikzlibrary{positioning}
\usepackage{subcaption}
\usepackage{mathrsfs}

\usepackage{amsthm}

\newtheorem{thm}{Theorem}

\newtheorem{lmm}{Lemma}
\newtheorem{dfn}{Definition}
\newtheorem{asm}{Assumption}

\newcommand{\thistheoremname}{}
\newtheorem*{genericthm*}{\thistheoremname}
\newenvironment{namedthm*}[1]
  {\renewcommand{\thistheoremname}{#1}%
  \begin{genericthm*}}
  {\end{genericthm*}}

\title{\LARGE \bf
How Does Driver Non-compliance Destroy Traffic Routing Control?
}

\author{Yu Tang, Li Jin and Kaan Ozbay
\thanks{This work was in part supported by US NSF Award CMMI-1949710, USDOT Award \# 69A3551747124 via the C2SMART Center, NYU Tandon School of Engineering, SJTU UM Joint Institute, and J. Wu \& J. Sun Endowment Fund.}
\thanks{Y. Tang is with C2SMART Center, Department of Civil \& Urban engineering, Tandon School of Engineering, New York University, 11201, USA.
        {\tt\small tangyu@nyu.edu}}%
\thanks{L. Jin is with UM Joint Institute and Department of Automation, Shanghai Jiao Tong University, Shanghai, 200240, China.
        {\tt\small li.jin@sjtu.edu.cn}}%
\thanks{K. Ozbay is with C2SMART Center, Department of Civil \& Urban engineering, Tandon School of Engineering, New York University, 11201, USA.
        {\tt\small kaan.ozbay@nyu.edu}}
}

\begin{document}

\maketitle
\thispagestyle{empty}
\pagestyle{empty}

\begin{abstract}
Routing control is one of important traffic management strategies against urban congestion. However, it could be compromised by heterogeneous drivers’ non-compliance with routing instructions. In this article we model the compliance in a stochastic manner and investigate its impacts on routing control. We consider traffic routing for two parallel links. Particularly, we consider two scenarios: one ignores congestion spillback while the other one considers it. We formulate the problem as a non-linear Markov chain, given random drivers’ adherence. Then we propose the stability and instability conditions to reveal when the routing is able or unable to stabilize the traffic. We show that for links without congestion spillback there exists a sufficient and necessary stability criterion. For links admiting congestion propagation, we present one stability condition and one instability condition. This stability conditions allow us to quantify the impacts of drivers' non-compliance on the two-link network in terms of throughput. Finally, we illustrate the results with a set of numerical examples.
\end{abstract}

\input{Sections/01_Introduction}
\input{Sections/02_Modeling}
\input{Sections/03_StabilityAnalysis}

\input{Sections/04_Example}
\input{Sections/05_Conclusion}




\bibliographystyle{IEEEtran}
\bibliography{Bibliography}

\end{document}

%% file: Sections/01_Introduction.tex
\section{INTRODUCTION}

\subsection{Motivation}
Dynamic traffic routing provides drivers with route recommendations based on real-time road information. It has been used as one of  promising control policies for alleviating congestion  \cite{como2012robust_2, minh2022effective}, and is expected to find extensive applications in a connected vehicle environment \cite{han2020congestion, chen2021distributed}. Nevertheless, it is also reported that driver non-compliance with route guidance could undermine the performance of dynamic routing \cite{powell2000value}, especially social routing advice that deliberately detours part of vehicles to achieve benefits in terms of road networks \cite{van2019travelers}. Although more and more surveys have confirmed this disobedience \cite{dia2007modelling,kerkman2012car,djavadian2014empirical,van2020travelers, mariotte2021assessing}, limited studies have investigated in an analytical way how drivers' adherence influences the effect of traffic routing control.

\begin{figure}[htbp]
    \centering
    \begin{tikzpicture}
        \Vertex[x=-3,style={color=white}]{A}
        \Vertex[x=0, style=black, size=0.1]{B}
        \Vertex[x=4, style=black, size=0.1]{N}
        \Edge[Direct, label={demand $D(t)$}](A)(B)
        \Edge[Direct,bend=30, label={major link $e_1$}](B)(N)
        \Edge[Direct,bend=330, label={minor link $e_2$}](B)(N)
        
        \Text[x=-0.3,y=-0.4, fontsize=\footnotesize]{Origin}
        \Text[x=4.3,y=-0.4, fontsize=\footnotesize]{Destination}
    \end{tikzpicture}
    \caption{The two-link network.}
    \label{fig_twolink}
\end{figure}
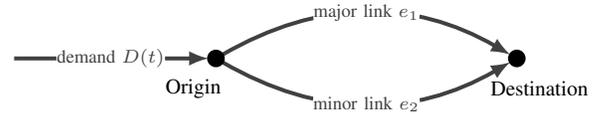

In this paper, we focus on routing advice released by traffic system operators/agencies. We study the above problem by considering a setting with random demand and random driver non-compliance. We analyze the resulting stochastic dynamical system under  routing control. Specially, we focus on a network comprised of two parallel links; see Fig.~\ref{fig_twolink}. Though simple, the two-link network serves as a typical scenario for studying routing control \cite{ephremides1980simple, zhang2019modeling,xie2020resilience,tang2020security}; it turns out to be an appropriate abstraction of multiple parallel links: one stands for arterials and the other denotes a set of local streets \cite{pi2017stochastic}. Furthermore, we adopt a Markov chain to model the compliance rate that possibly depends on traffic states. It allows us to study stability and instability criteria that determine whether the network is destabilized by the random compliance rate. We also quantify the impacts of drivers' disobedience in terms of throughput, namely the maximum constant inflow under which the network can be stabilized.

\subsection{Related work}

Previous work on evaluating impacts of the conformity with routing advice typically applied static or dynamic traffic assignment (STA or DTA). These methods are favored since they easily provide numerical assessment in terms of efficiency, equity and so on \cite{van2019travelers, zhang2019modeling,ning2023robust} and can be implemented even for large-scale networks.  However, they also have disadvantages. STA finds equilibrium by solving mathematical programming. It fails to capture significant traffic dynamics, such as congestion spillback and fluctuations of drivers' compliance rate and thus could induce unrealistic equilibrium. Though DTA can address the shortcomings of STA to some extent, it introduces a new problem. As we see later, low compliance rates could make traffic networks unstable. In that case, it could be problematic to apply DTA since we do not have guaranteed convergence in advance. Noting this, we aim at developing methods that allow stability and instability analysis, at least in some conditions, before numerical evaluation. To our best knowledge, limited studies discussed this topic for routing control subject to random compliance.




Our model belongs to discrete-time nonlinear stochastic systems. Although the general theories of stochastic stability have been studied extensively \cite{berman2006h,zhang2016lasalle, zhang2021robust,meyn1993survey,meyn2012markov}, how to apply them in our problem is still unclear. Typically, stability analysis can be refined for specific non-linear systems. Besides, it should be noted that most of studies mainly discuss sufficient stability conditions for general nonlinear stochastic systems, while we also have interest in instability conditions.

\subsection{Our contributions}

In this paper we address the two following questions:
\begin{enumerate}
    \item How to determine whether the network can be stabilized by routing control subject to driver non-compliance?
    \item How to evaluate efficiency losses of routing control due to the non-compliance in an analytic way?
\end{enumerate}

We answer the first question for two types of networks. In the first one, the two parallel links have infinite space and there are no congestion spillback, while in the second one, the two parallel links with finite space may. We formulate discrete-time nonlinear stochastic systems for the two networks, respectively. Then we apply the Foster-Lyapunov criteria \cite{meyn2012markov} to derive the stability condition and scrutinize transience of Markov chains \cite{meyn2012markov,meyn1993survey} to obtain instability conditions. For the first network, we successfully obtain a sufficient and necessary stability condition (Theorem~\ref{thm_1}); for the second one, we have one stability crietrion (Theorems \ref{thm_2}) and one instability criterion (Theorems \ref{thm_3}). 

Even when the network is stable, we want to know to what extent the network performance decrease. Thus to answer the second question, we take throughput as the metric to measure efficiency losses. However, throughput is not always available even for the two-link network. For the two links with infinite buffer sizes, we indeed derive exact values of throughput since we have a sufficient and necessary stability condition. For the two links with finite buffer sizes, we use the stability and instability conditions to yield lower and upper bounds, respectively.  

The rest of the paper is organized as follows. Section~\ref{sec_modeling} introduces our modeling framework.
Section~\ref{sec_stability} presents the results when the two parallel links have infinite storage space, and Section~\ref{sec_example} provides the results in case of two links with limited buffer sizes. Finally, Section~\ref{sec_conclusion} summarizes our work and discusses future research.


%% file: Sections/02_Modeling.tex
\section{Modeling and formulation}
\label{sec_modeling}

Consider the two-link network in Fig.~\ref{fig_twolink}: one is the major link $e_1$, typically with a higher free-flow speed or capacity, and the other is the minor link $e_2$. We suppose that the system operator tries to route part of flows to the minor link $e_2$ to reduce congestion in the major link $e_1$.

We denote by $X_e(t)\in\mathbb{R}_{\geq0}$ traffic density of link $e\in\{e_1,e_2\}$ at time $t$. Each link $e\in\{e_1, e_2\}$ is associated with a sending flow $f_e(x_e):\mathbb{R}_{\geq0}\to\mathbb{R}_{\geq0}$ and a receiving flow $r_e(x_e):\mathbb{R}_{\geq0}\to\mathbb{R}_{\geq0}$. Here the sending flow $f_e(x_e)$ indicates the desired outflow from link $e$ given traffic density $x_e$, and the receiving flow $r_e(x_e)$ stands for the maximum flow allowed into link $e$. We assume that the flow functions satisfy:
\begin{asm}[Sending \& receiving flows]
\label{asm_1}
\quad

\begin{enumerate}
    \item[1.1] Sending flows: For link $e$, $f_e(x_e)$ is Lipschitz continuous and $\mathrm{d}f_e(x_e)/\mathrm{d}x_e\geq0$ almost everywhere (a.e.). Moreover, $f_e(0)=0$ and $\sup_{x_e}f_e(x_e)<\infty$.
    \item[1.2] Receiving flows: For link $e$ with a finite buffer size $x_e^{\max}<\infty$, $r_e(x_e)$ is Lipschitz continuous and $\mathrm{d}r_e(x_e)/\mathrm{d}x_e\leq0$ a.e.. Moreover, $r_e(x_e^{\max})=0$ and $\sup_{x_e}r_e(x_e)<\infty$. For link $e$ with an infinite buffer size, $r_e=\infty$. 
\end{enumerate}
\end{asm}

The assumptions above follow the conventional modeling of road traffic. We also define \emph{link capacity} as
\begin{equation}
    Q_e := \sup_{x_e} \min\{f_e(x_e), r_e(x_e)\},
\end{equation}
which denotes an upper bound of sustainable discharging flow from link $e$.

Note that Assumption~\ref{asm_1}.2 implies that it is reasonable to only consider $X_e(t)\in[0, x_e^{\max}]$ for link $e$ with limited storage. Compared with supposing finite buffer sizes, the assumption of infinite buffer sizes seems a little unrealistic, but it helps understand and design routing control, even on complex networks. In this paper, we discuss both of them.

For demand modeling, we consider an independent and identically distributed (i.i.d.) stochastic process $\{D(t):t\geq0\}$ with a distribution $\Gamma^d$, $\mathbb{E}[D(t)]=\alpha$ and $D(t)\in\mathcal{D}$ for $t\geq 0$, where $\mathcal{D}$ is a compact set. This is based on the observation that during rush hours, of interest to traffic management, traveling demands are relatively stationary and only fluctuate within certain bounds \cite{adot}. Obviously, we require 
\begin{equation}
    \mathbb{E}[D(t)]=\alpha < Q_{e_1} + Q_{e_2}, \label{eq_necess}
\end{equation}
otherwise the traffic densities must blow up.

Next, we introduce routing control. Let $\beta_e(x):\mathbb{R}^2_{\geq0}\to[0,1]$ denote a proportion of traffic routed to link $e$. We assume the routing policies to satisfy:
\begin{asm}[Routing control] 
The routing proportions $\beta_{e_1}(x_{e_1}, x_{e_2})$ and $\beta_{e_2}(x_{e_1}, x_{e_2})$ are continuous and have the following monotonicity a.e.:
\begin{enumerate}
    \item[2.1] $\frac{\partial}{\partial x_{e_1}}\beta_{e_1}(x_{e_1},x_{e_2}) \leq 0$ and  $\frac{\partial}{\partial x_{e_2}}\beta_{e_1}(x_{e_1},x_{e_2}) \geq 0$;
    \item[2.2] $\frac{\partial}{\partial x_{e_1}}\beta_{e_2}(x_{e_1},x_{e_2}) \geq 0$ and  $\frac{\partial}{\partial x_{e_2}}\beta_{e_2}(x_{e_1},x_{e_2}) \leq 0$.
\end{enumerate}
\end{asm}

The assumption above implies that the routing proportion $\beta_{e}(x_{e_1},x_{e_2})$ tends to decrease (resp. increase) as link $e$ (resp. the other link) becomes more congested. It holds true for typical routing policies, such as logit routing \cite{como2011robust}.

Recalling that routing proportions could be compromised due to heterogeneous drivers' choice behavior, we denote by $C(t)\in[0, 1]$ the compliance rate of drivers' routed to the minor link $e_2$ at time $t$. Then the compromised routing ratios, denoted by $\tilde{\beta}_e(x_{e_1},x_{e_2},c):\mathbb{R}^2_{\geq0}\times[0, 1]\to[0,1]$, are given by
\begin{subequations}
    \begin{align}
        &\tilde{\beta}_{e_1}(X_{e_1}(t), X_{e_2}(t), C(t)) \nonumber \\
        &=\beta_{e_1}(X_{e_1}(t), X_{e_2}(t)) + \beta_{e_2}(X_{e_1}(t), X_{e_2}(t))(1-C(t)), \label{eq_comp_1} \\
        &\tilde{\beta}_{e_2}(X_{e_1}(t),X_{e_2}(t), C(t)) \nonumber \\
        &= \beta_{e_2}(X_{e_1}(t), X_{e_2}(t))C(t). \label{eq_comp_2}
    \end{align}
\end{subequations}
Note that \eqref{eq_comp_1}-\eqref{eq_comp_2} imply the compliance rate of drivers routed to the major link $e_1$ equals one. This is because in our setting drivers are assumed to prefer the major link $e_1$ while the system operator tries to route some of them to the minor link $e_2$. The assumption is not necessary, just for simplifying the problem. In fact, we can introduce the second compliance rate, and apply our method to obtain stability and instability criteria, which are more complicated.

We consider that $C(t+1)$ depends on $X_{e_1}(t)=x_{e_1}$ and $X_{e_2}(t)=x_{e_2}$ with a distribution $\Gamma^c_{x_{e_1},x_{e_2}}$. For convenience of analysis, we assume that the distributions $\Gamma^c_{x_{e_1},x_{e_2}}(c)$, for any $x_{e_1}$ and $x_{e_2}$, have lower semi-continuous densities with the same support $\mathcal{C}\subseteq [0,1]$. We define $\mathbb{E}_{x_{e_1},x_{e_2}}[C]:=\mathbb{E}[C(t+1)|X_{e_1}(t)=x_{e_1},X_{e_2}(t)=x_{e_2}]$ and assume it to satisfy:
\begin{asm}[Drivers' compliance] The expected compliance rate has the following monotonicity a.e.:
\begin{equation}
    \frac{\partial}{\partial x_{e_1}} \mathbb{E}_{x_{e_1},x_{e_2}}[C] \geq0\text{, and } \frac{\partial}{\partial x_{e_2}} \mathbb{E}_{x_{e_1},x_{e_2}}[C] \leq0. 
\end{equation}
\end{asm}

Clearly, the assumption implies that more drivers follow the routing advise to the minor link $e_2$ if the major link $e_1$ becomes more congested or the minor link $e_2$ becomes less congested.

The following specifies the inflows into links $e_1$ and $e_2$. Given an upstream flow $F(t)$, we denote by $q_e^{\mathrm{in}}:\mathbb{R}_{\geq0}^3\times[0,1]\to\mathbb{R}_{\geq0}$ the inflow into link $e\in\{e_1,e_2\}$:
\begin{align}
&q_e^{\mathrm{in}}(F(t), X_{e_1}(t), X_{e_2}(t), C(t)) \nonumber \\
&= \min\{\tilde{\beta}_e(X_{e_1}(t), X_{e_2}(t), C(t)) F(t), r_{e}(X_{e}(t))\}.
\end{align}

Supposing $r_{e_1}=r_{e_2}=\infty$, we have the following network dynamics:
\begin{subequations}
    \begin{align}
        \Delta X_{e_1}(t) =& \frac{\delta}{l_{e_1}}\Big(q_{e_1}^{\mathrm{in}}(D(t), X_{e_1}(t), X_{e_2}(t), C(t)) \nonumber \\ 
        & - f_{e_1}(X_{e_1}(t))\Big), \label{eq_inf_1} \\
        \Delta X_{e_2}(t) =& \frac{\delta}{l_{e_2}}\Big(q_{e_2}^{\mathrm{in}}(D(t), X_{e_1}(t), X_{e_2}(t), C(t)) \nonumber \\ 
        & - f_{e_2}(X_{e_2}(t))\Big). \label{eq_inf_2}
    \end{align}
\end{subequations}
where $\Delta X_e(t):=X_e(t+1)-X_e(t)$ for any link $e$, $\delta$ denotes the time step size and $l_e$ denotes length of link $e$.

Clearly, if links $e_1$ and $e_2$ have finite space, congestion could block the inflows. For the sake of analysis, we consider another link $e_0$ upstream of links $e_1$ and $e_2$, satisfying $Q_{e_0}\geq Q_{e_1}+Q_{e_2}$ and $r_{e_0}=\infty$, to accept inflows. It leads to the network dynamics as follows:
\begin{subequations}
    \begin{align}
        &\Delta X_{e_0}(t) = \frac{\delta}{l_{e_0}}\Big(D(t) \nonumber \\
        &-\sum_{e\in\{e_1,e_2\}} q_e^{\mathrm{in}}(f_{e_0}(X_{e_0}(t)), X_{e_1}(t), X_{e_2}(t), C(t)) \Big) \label{eq_fin_1} \\
        &\Delta X_{e_1}(t) = \frac{\delta}{l_{e_1}}\Big(q_{e_1}^{\mathrm{in}}(f_{e_0}(X_{e_0}(t)), X_{e_1}(t), X_{e_2}(t), C(t)) \nonumber \\
        & - f_{e_1}(X_{e_1}(t))\Big), \label{eq_fin_2}  \\
        &\Delta X_{e_2}(t) = \frac{\delta}{l_{e_2}}\Big(q_{e_2}^{\mathrm{in}}(f_{e_0}(X_{e_0}(t)), X_{e_1}(t), X_{e_2}(t), C(t)) \nonumber \\
        &-f_{e_2}(X_{e_2}(t))\Big). \label{eq_fin_3} 
    \end{align}
\end{subequations}
For notional convenience, we assume $\delta/l_e$  to be the same for any link $e$ and ignore them in the following analysis.

Then, \eqref{eq_inf_1}-\eqref{eq_inf_2} indicate that
\begin{equation}
    \Phi_1:=\{(X_{e_1}(t), X_{e_2}(t), D(t), C(t)):t\geq0\} \label{eq_markovchain_1} 
\end{equation}
is a Markov chain with a state space $\mathbb{R}_{\geq0}\times\mathbb{R}_{\geq0}\times\mathcal{D}\times\mathcal{C}$, and \eqref{eq_fin_1}-\eqref{eq_fin_3} indicate 
\begin{equation}
    \Phi_2:=\{(X_{e_0}(t), X_{e_1}(t), X_{e_2}(t), D(t), C(t)):t\geq0\} \label{eq_markovchain_2}
\end{equation}
is also a Markov chain with a state space $\mathbb{R}_{\geq0}\times\mathcal{X}_{e_1}\times\mathcal{X}_{e_2}\times\mathcal{D}\times\mathcal{C}$. Note that $\mathcal{X}_{e_1}\subseteq[0, x_{e_1}^{\max}]$ and $\mathcal{X}_{e_2}\subseteq[0, x_{e_2}^{\max}]$ are bounded sets.

We make the last assumption as follows:
\begin{asm}
\quad

\begin{enumerate}
    \item[4.1] For the system \eqref{eq_inf_1}-\eqref{eq_inf_2}, there exists $c\in\mathcal{C}$ and $d\in\mathcal{D}$ such that $\lim_{t\to\infty} X_e(t)=x_e^*<\infty$, $e\in\{e_1,e_2\}$, given $C(t)\equiv c$ and $D(t)\equiv d$;
    \item[4.2] For the system \eqref{eq_fin_1}-\eqref{eq_fin_3}, there exists $c\in\mathcal{C}$ and $d\in\mathcal{D}$ such that $\lim_{t\to\infty} X_e(t)=x_e^*<\infty$, $e\in\{e_0,e_1,e_2\}$, given $C(t)\equiv c$ and $D(t)\equiv d$. Moreover, 
    \begin{subequations}
        \begin{align}
            \tilde{\beta}_{e_1}((x_{e_1}^*, x_{e_2}^*), c)f_{e_0}(x_{e_0}^*) <& r_{e_1}(x_{e_1}^*), \label{eq_asm4_1} \\
            \tilde{\beta}_{e_2}((x_{e_1}^*, x_{e_2}^*), c)f_{e_0}(x_{e_0}^*) <& r_{e_2}(x_{e_2}^*). \label{eq_asm4_2} 
        \end{align}
    \end{subequations}
\end{enumerate}
\end{asm}

The above assumption essentially states that there exists $c$ and $d$ such that the systems \eqref{eq_inf_1}-\eqref{eq_inf_2} and \eqref{eq_fin_1}-\eqref{eq_fin_3} are stable. Note that \eqref{eq_asm4_1}-\eqref{eq_asm4_2} are mild technical assumptions. The system \eqref{eq_inf_1}-\eqref{eq_inf_2} does not require them due to $r_{e_1}=r_{e_2}=\infty$. The equations \eqref{eq_asm4_1}-\eqref{eq_asm4_2} imply that the inflows into links $e_1$ and $e_2$ are strictly fewer than the corresponding receiving flows. That is, the inflow can smoothly pass links $e_1$ and $e_2$ when there is no congestion. By noting \eqref{eq_necess},  \eqref{eq_asm4_1}-\eqref{eq_asm4_2} are easy to achieve for appropriate routing polices.

We have the following lemma proved in Appendix~\ref{app_pf_lmm1}:
\begin{lmm} 
\label{lmm_1}
Given Assumption~4.1, the Markov chain \eqref{eq_markovchain_1} is $\varphi$-irreducible; and given Assumption~4.2, the Markov chain \eqref{eq_markovchain_2} is $\varphi$-irreducible.
\end{lmm}

Here $\varphi$ is a certain measure. The $\varphi$-irreducibility means that any set with positive measure can be reached by the Markov chain given any initial state. It implies that any large set can be reached from any initial condition and thus the state space is indecomposable. It is a prerequisite of discussing stability of Markov chains.

Finally, we define the stability of interest below:
\begin{dfn}[Stability \& Instability]
A stochastic process $\{Y(t):t\geq0\}$ with a state space $\mathcal{Y}$ is \emph{stable} if there exists a scalar $Z<\infty$ such that for any initial condition $y\in\mathcal{Y}$
\begin{equation}\label{eq_bounded}
  \limsup_{t\to\infty}\frac{1}{t}\sum_{\tau=0}^t\mathbb {E}[|Y(\tau)|] \le Z,
\end{equation}
where $|Y(\tau)|$ denotes 1-norm of $Y(\tau)$. The network is \emph{unstable} if there does not exist $Z<\infty$ such that \eqref{eq_bounded} holds for any initial condition $y\in\mathcal{Y}$.
\end{dfn}

The notion of stability follows a classical definition \cite{dai1995stability} and is widely used in studying traffic control \cite{barman2023throughput}. Practically, if the time-average traffic density in all links are bounded, the network is stable; otherwise, it is unstable.



%% file: Sections/03_StabilityAnalysis.tex
\section{Stability analysis of the network without congestion propagation}
\label{sec_stability}

We state the main result as follows:
\begin{thm}
\label{thm_1}
The Markov chain \eqref{eq_markovchain_1} with the state space $\mathbb{R}_{\geq0}\times\mathbb{R}_{\geq0}\times\mathcal{D}\times\mathcal{C}$ is stable if and only if there exists a vector $\theta:=[\theta_{e_1},\theta_{e_2}]^{\mathrm{T}}\in\mathbb{R}_{\geq0}^2$ such that
\begin{subequations}
    \begin{align}
        \Big(\beta_{e_1}(\theta) + \beta_{e_2}(\theta)\mathbb{E}_\theta[1-C]\Big)\alpha -f_{e_1}(\theta_{e_1}) &< 0, \label{eq_thm1_1} \\
        \beta_{e_2}(\theta)\mathbb{E}_\theta[C]\alpha  -f_{e_2}(\theta_{e_2}) &< 0. \label{eq_thm1_2}
    \end{align}
\end{subequations}
\end{thm}

Note that the stability condition is sufficient and necessary. Thus we can use it to derive exact values of throughput. In the following sections, we first present a numerical example and then prove Theorem~\ref{thm_1}.

\subsection{Numerical example}
\label{sec_num_inf}

\begin{figure}[htbp]
    \centering
    \includegraphics[width=0.6\linewidth]{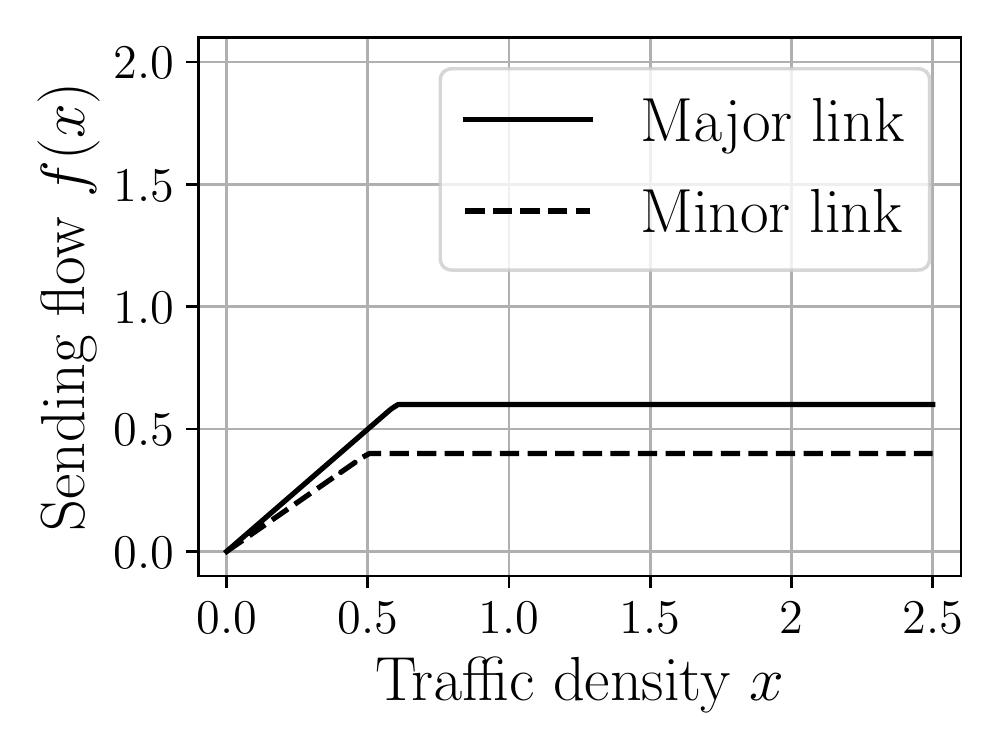}
    \caption{Sending flows of major and minor links.}
    \label{fig_sendingflow}
\end{figure}

First, we set $\delta = 0.1$ and  $l_{e_1}=l_{e_2}=1$. We consider the sending flows 
\begin{equation}
    f_e(x_e) = \min\{v_ex_e, Q_e\}, ~e\in\{e_1,e_2\} \label{eq_sending}
\end{equation}
with $v_{e_1}=1$, $v_{e_2}=0.8$, $Q_{e_1}=0.6$, $Q_{e_2}=0.4$, as illustrated in Fig.~\ref{fig_sendingflow}.

For the purpose of routing, we adopt the classical logit routing as follows:
\begin{equation}
     \beta_e(x) = \frac{e^{-\nu_e x_e}}{e^{-\nu_{e_1}x_{e_1}}+e^{-\nu_{e_2}x_{e_2}}}, ~e\in\{e_1,e_2\},
\end{equation}
where $\nu_{e_1}=1$ and $\nu_{e_2}=2$ are routing parameters. 

We assume the demands $D(t)\in[\underline{d}, 1.2]$, $t\geq0$, are independent and identically distributed (i.i.d.) uniform random variables. It follows $\mathbb{E}[D(t)]=\underline{d}/2+0.6$.  We also assume the routing compliance rates $C(t)\in[0,\bar{c}]$, $t\geq0$, are i.i.d. uniform random variables, along with $\mathbb{E}[C(t)]=\bar{c}/2$. It indicates that in our numerical example the compliance rates are independent of traffic states. It should be noted that this independence is not necessary for our approach. Here we assume it just for simplification. However, we still have non-trivial observations in this case.

\begin{figure}[htbp]
    \centering
    \begin{subfigure}{0.45\linewidth}
    \centering
    \includegraphics[width=\linewidth]{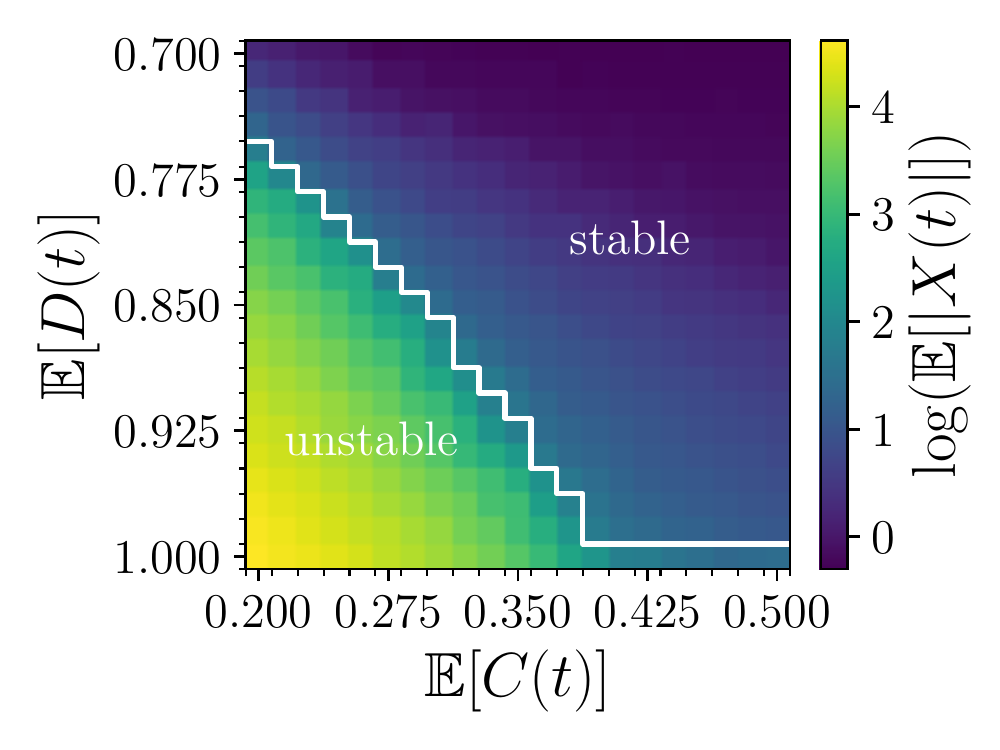}  
  \caption{Stability region.}
  \label{fig_region_inf}
\end{subfigure}
\begin{subfigure}{0.45\linewidth}
  \centering
  \includegraphics[width=\linewidth]{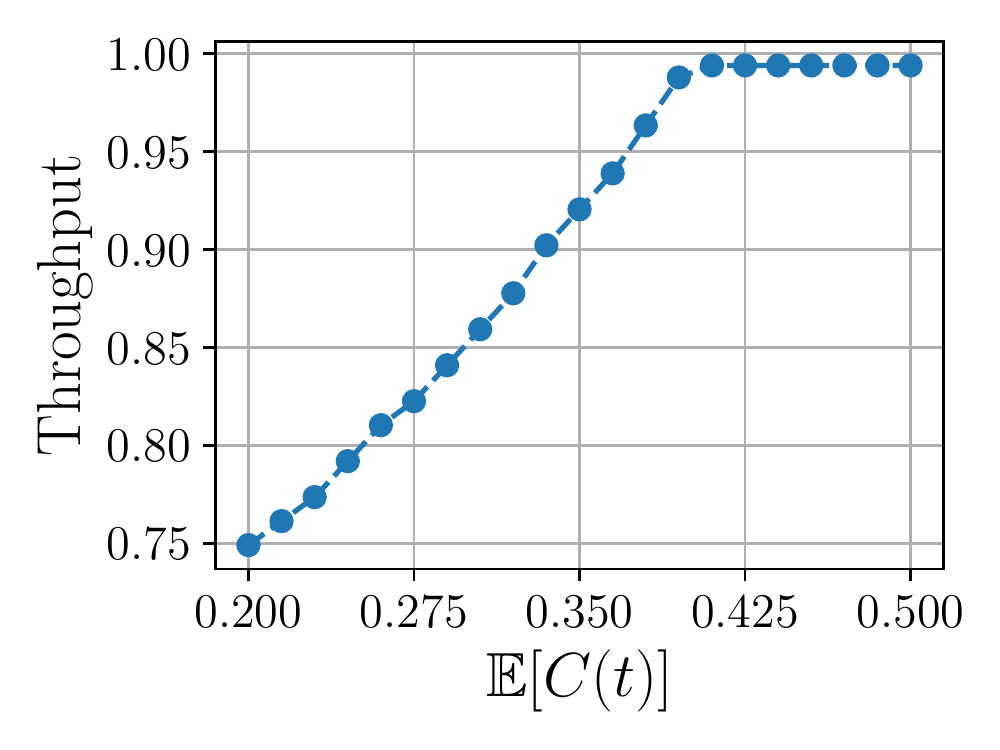}  
  \caption{Throughput.}
  \label{fig_throu_inf}
\end{subfigure}
    \caption{Analysis of stability and throughput given links $e_1$ and $e_2$ with infinite buffer sizes.}
    \label{fig_infinite_analysis}
\end{figure}

We first analyze the stability and instability of scenarios with different $\underline d$ and compliance rate $\bar c$. Fig.~\ref{fig_region_inf} shows the time-average traffic densities after $5\times10^5$ steps and reveals the stability and instability regions. We observe a non-linear boundary: given moderate traffic demands, improvements of compliance rates can stabilize the network; but given a high demand close to the network capacity, we hardly see the effect of improving compliance rate.

Then we compute the throughput, the maximum expected demand under which the network can be stabilized. It is interesting to find that we can achieve a relatively high throughput (around 0.987) when $\mathbb{E}[C(t)]=0.395$. Further improvement is marginal when $\mathbb{E}[C(t)]$ exceed 0.395.  

\subsection{Proof of Theorem~\ref{thm_1}}
We first prove the sufficiency by the Foster-Lyapunov criterion \cite{meyn2012markov}:
\begin{namedthm*}{Foster-Lyapunov criterion}
Consider a $\varphi$-irreducible Markov chain $\{Y(t); t\ge0\}$ with a state space $\mathcal Y$, an infinitesimal generator $\mathscr L$, and a Lyapunov function $V:\mathcal Y\to\mathbb R_{\ge0}$. If there exist constants $m>0$, $n<\infty$, a function $g:\mathcal Y\to\mathbb R_{\ge0}$ and a compact set $\mathcal{E}$ such that for any $y\in\mathcal{Y}$
\begin{align*}
    \mathbb{E}[V(Y(t+1))|Y(t)=y] - V(y) \leq -mg(x) + n\mathbf{1}_{\mathcal{E}}(y),
\end{align*}
where $\mathbf{1}_{\mathcal{E}}(y)$ is an indicator function, then, for each initial condition $y(0)\in\mathcal Y$,
$$
\limsup_{t\to\infty}\frac1t\sum_{\tau=0}^t\mathrm E[g(Y(\tau))] \le m/n.
$$
\end{namedthm*}

To proceed, we consider the following Lyapunov function
\begin{equation}
    V(x) = \begin{cases}
    0 & x\in\mathcal{X}^1, \\
    \frac{1}{2}(x_{e_1}-\theta_{e_1})_+^2 & x\in\mathcal{X}^2, \\
    \frac{1}{2}(x_{e_2}-\theta_{e_2})_+^2 & x\in\mathcal{X}^3, \\
    \frac{1}{2}((x_{e_1}-\theta_{e_1})_+ + (x_{e_2}-\theta_{e_2})_+)^2 & x\in\mathcal{X}^4,
    \end{cases}
\end{equation}
where $(\cdot)_+:=\max\{\cdot, 0\}$, $\mathcal{X}_{e_1}:=[0,\theta_{e_1}]\times[0,\theta_{e_2}]$, $\mathcal{X}_{e_2}:=(\theta_{e_1}, \infty)\times[0,\theta_{e_2}]$, $\mathcal{X}_{e_3}:=[0,\theta_{e_1}]\times(\theta_{e_2}, \infty)$ and $\mathcal{X}_{e_4}:=(\theta_{e_1}, \infty)\times(\theta_{e_2}, \infty)$.

The rest is devoted to show that there exist constants $m'>0$ and $n'<\infty$ such that for every $x\in\mathbb{R}_{\geq0}^2$
\begin{align}
 &\mathbb{E}[V(X(t+1))|X(t)=x] - V(x) \nonumber \\
 \leq & -m' \Big((x_{e_1}-\theta_{e_1})_+ + (x_{e_2}-\theta_{e_2})_+\Big) + n'.  \label{eq_proof_thm1_1}  
\end{align}
If \eqref{eq_proof_thm1_1} holds, we must have $0<m<m'$, $n<\infty$, and a compact set $\mathcal{E}=[0, M]\times[0,M]$ such that
\begin{align}
 &\mathbb{E}[V(X(t+1))|X(t)=x] - V(x) \nonumber \\
 \leq & -m \Big((x_{e_1}-\theta_{e_1})_+ + (x_{e_2}-\theta_{e_2})_+\Big) + n\mathbf{1}_{\mathcal{E}}(x),  \label{eq_proof_thm1_2}  
\end{align}
which indicates that $X_{e_1}(t)$ and $X_{e_2}(t)$ and thus concludes the stability.

To show \eqref{eq_proof_thm1_1}, we need to discuss whether $X_{e_1}(t)$ is larger than $\theta_{e_1}$ and whether $X_{e_2}(t)$ is larger than $\theta_{e_2}$, up to four cases. Here we present the proofs for the typical cases and the remaining can be proved in a similar way.

When $X_{e_1}(t)\leq\theta_{e_1}$ and $X_{e_2}(t)\leq\theta_{e_2}$, the proof is trivial by noting that $X_{e_1}(t+1)$ and $X_{e_2}(t+1)$ must be bounded a sufficiently large number.

Now we assume $X_{e_1}(t)>\theta_{e_1}$ and $X_{e_2}(t)\leq\theta_{e_2}$. It follows
\begin{align*}
    &\mathbb{E}[V(X(t+1))|X(t)=x]-V(x) \\
    \leq& \frac{1}{2}\Big(\int (x_{e_1}+G_{e_1}-\theta_{e_1})_+^2 - (x_{e_1}-\theta_{e_1})^2\Big)  \\
    \leq& \frac{1}{2}\Big(\int (x_{e_1}+G_{e_1}-\theta_{e_1})^2 - (x_{e_1}-\theta_{e_1})^2\Big) \\
    \leq& \Big(\Big(\beta_{e_1}(x) + \beta_{e_2}(x)\mathbb{E}_x[1-C]\Big)\alpha -f_{e_1}(x_{e_1}) \Big)x_{e_1} + n,
\end{align*}
where $n$ is a sufficiently large number and
\begin{equation}
    G_{e_1}:=q_{e_1}^{\mathrm{in}}(D(t), X_{e_1}(t), X_{e_2}(t), C(t)) - f_{e_1}(X_{e_1}(t)). \label{eq_G_e1}
\end{equation}
Note that we omit $\delta/l_{e_1}$. By Assumptions~1-3,
$$\Big(\beta_{e_1}(x_{e_1}, x_{e_2}) + \beta_{e_2}(x_{e_1}, x_{e_2})\mathbb{E}_x[1-C]\Big)\alpha -f_{e_1}(x_{e_1})$$
is non-increasing in $x_{e_1}$ and non-decreasing in $x_{e_2}$. Thus \eqref{eq_thm1_1} indicates that there exists $m_1'>0$ such that for any $x_{e_1}>\theta_{e_1}$ and $x_{e_2}\leq\theta_{e_2}$,
\begin{align*}
    &\mathbb{E}[V(X(t+1))|X(t)=x]-V(x) \\
    \leq& -m_1'x_{e_1}+n \\
    =& -m_1'\Big((x_{e_1}-\theta_{e_1})_++(x_{e_2}-\theta_{e_2})_+\Big) + (n-m_1'\theta_{e_1}).
\end{align*}
Since $n$ can be sufficiently large, we must have 
$$n-m_1'\theta_{e_1}>0.$$

Finally, we consider $X_{e_1}(t)>\theta_{e_1}$ and $X_{e_2}(t)>\theta_{e_2}$. It turns out that we obtain
\begin{align}
    &\mathbb{E}[V(X(t+1))|X(t)=x]-V(x) \nonumber \\
    \leq & \Big(\alpha - f_{e_1}(x_{e_1}) - f_{e_2}(x_{e_2})\Big)(x_{e_1}+x_{e_2}) + n. \nonumber
\end{align}
Note the $\alpha - f_{e_1}(x_{e_1})) - f_{e_2}(x_{e_2}$ is non-increasing in both $x_{e_1}$ and $x_{e_2}$. Combing \eqref{eq_thm1_1}-\eqref{eq_thm1_2} indicates that there exists $m_2'>0$ such that for any $x_{e_1}>\theta_{e_1}$ and $x_{e_2}>\theta_{e_2}$,
\begin{align*}
    &\mathbb{E}[V(X(t+1))|X(t)=x]-V(x) \\
    \leq& -m_2'(x_{e_1}+x_{e_2}) + n \\
    =& -m_2'\Big((x_{e_1}-\theta_{e_1})_++(x_{e_2}-\theta_{e_2})_+\Big) \\
    & + (n-m_2'\theta_{e_1}-m_2'\theta_{e_2}).
\end{align*}
We can similarly conclude that $n-m_2'\theta_{e_1}-m_2'\theta_{e_2}>0$. Thus we finish proving the sufficient condition.

The following proves the necessary condition. That is, we show that if there does not exist a vector $\theta$ satisfying \eqref{eq_thm1_1}-\eqref{eq_thm1_2}, the system is unstable. We prove the instability by showing the Markov chain \eqref{eq_markovchain_1} is transient, which indicates that some state tends towards infinity in the long term, and thus transience can be as instability \cite{meyn1993survey}. We consider the following transience criterion \cite{meyn2012markov}:
\begin{namedthm*}{Transience criterion}
Consider a $\varphi$-irreducible Markov chain $\{Y(t): t\ge0\}$ with a state space $\mathcal Y$. Then $\{Y(t): t\geq0\}$ is transient if there exists a bounded function $V:\mathcal{Y}\to\mathbb{R}_{\geq0}$ and a sublevel set of $V$, denoted by $S$, such that 
\begin{enumerate}
    \item[(i)] $\varphi(S)>0$ and $ \varphi(\mathcal{Y}\setminus S)>0$;
    \item[(ii)] $\mathbb{E}[V(Y(t+1))|Y(t)=y]-V(y)\geq0,~\forall y\in \mathcal{Y}\setminus S$.
\end{enumerate}
\end{namedthm*}

Note that our Markov chain \eqref{eq_markovchain_1} is $\varphi$-irreducible, stated by Lemma~\ref{lmm_1}. To proceed, we first assume that for any $\theta\in\mathbb{R}_{\geq0}^2$
\begin{equation}
    \Big(\beta_{e_1}(\theta) + \beta_{e_2}(\theta)\mathbb{E}_\theta[1-C]\Big)\alpha -f_{e_1}(\theta_{e_1}) \geq 0. \label{eq_pf_thm1_transience}
\end{equation}
We consider a bounded test function $W:\mathbb{R}_{\geq0}\to\mathbb{R}_{\geq0}$:
\begin{equation}
    W(x_{e_1}) = \xi_1 - \frac{1}{x_{e_1} + \xi_2}
\end{equation}
where $\xi_1$ and $\xi_2$ are sufficiently large numbers.
Then, we obtain
\begin{align*}
     &\mathbb{E}[W(X_{e_1}(t+1))|X_{e_1}(t)=x_{e_1}] - f_{e_1}(x_{e_1}) \\
    =& \frac{1}{x_{e_1}+\xi_2} - \int \frac{1}{x_{e_1}+G_{e_1}+\xi_2} \\
    =& \int\frac{G_{e_1}}{(x_{e_1}+\xi_2)(x_{e_1}+G_{e_1}+\xi_2)},
\end{align*}
where $G_{e_1}$ is given by \eqref{eq_G_e1}. Note that $G_{e_1}$ is bounded because flow functions are bounded. With loss of generality, we assume 
\begin{equation*}
    G_{e_1}^{\mathrm{min}} \leq G_{e_1} \leq G_{e_1}^{\mathrm{max}}.
\end{equation*}
If follows
\begin{align*}
     &\mathbb{E}[W(X_{e_1}(t+1))|X_{e_1}(t)=x_{e_1}] - f_{e_1}(x_{e_1}) \\
    =& \int_{G_{e_1}\leq0}\frac{G_{e_1}}{(x_{e_1}+\xi_2)(x_{e_1}+G_{e_1}+\xi_2)} \\
     & + \int_{G_{e_1}>0}\frac{G_{e_1}}{(x_{e_1}+\xi_2)(x_{e_1}+G_{e_1}+\xi_2)} \\
    \geq & \int_{G_{e_1}\leq0}\frac{G_{e_1}}{(x_{e_1}+\xi_2)(x_{e_1}+G_{e_1}^{\mathrm{min}}+\xi_2)} \\
     & + \int_{G_{e_1}>0}\frac{G_{e_1}}{(x_{e_1}+\xi_2)(x_{e_1}+G_{e_1}^{\mathrm{max}}+\xi_2)} \\
     =& \frac{1}{(x_{e_1}+\xi_2)(x_{e_1}+G_{e_1}^{\mathrm{min}}+\xi_2)}\int_{G_{e_1}\leq0} G_{e_1} \\
     &+ \frac{1}{(x_{e_1}+\xi_2)(x_{e_1}+G_{e_1}^{\mathrm{max}}+\xi_2)}\int_{G_{e_1}>0} G_{e_1} \\
     =& \frac{1}{(x_{e_1}+\xi_2)(x_{e_1}+G_{e_1}^{\mathrm{min}}+\xi_2)(x_{e_1}+G_{e_1}^{\mathrm{max}}+\xi_2)}\Big(\\
     &(x_{e_1}+\xi_2)\int G_{e_1}+ G_{e_1}^{\max}\int_{G_{e_1}\leq0}G_{e_1} \\
     & + G_{e_1}^{\min}\int_{G_{e_1}>0}G_{e_1} \Big).
\end{align*}
Obviously, given 
$$\int G_{e_1} = \Big(\beta_{e_1}(\theta) + \beta_{e_2}(\theta)\mathbb{E}_\theta[1-C]\Big)\alpha -f_{e_1}(\theta_{e_1}) \geq 0,$$
we must have 
$$\mathbb{E}[W(X_{e_1}(t+1))|X_{e_1}(t)=x_{e_1}] - f_{e_1}(x_{e_1})\geq0$$
by letting $\xi_2$ be sufficiently large.

Note that the above inequality holds over $\mathbb{R}_{\geq0}^2$. It indicates that the conditions (i)-(ii) above are satisfied. Thus we conclude the Markov chain \eqref{eq_markovchain_1} is transient given \eqref{eq_pf_thm1_transience}.

Then we assume that for any $\theta\in\mathbb{R}_{\geq0}^2$
\begin{equation}
    \beta_{e_2}(\theta)\mathbb{E}_\theta[C]\alpha  -f_{e_2}(\theta_{e_2}) \geq 0.
\end{equation}
We can prove in a similar way that the Markov chain \eqref{eq_markovchain_1} is transient in this case. Finally, we conclude that if there does not exist a vector $\theta$ satisfying \eqref{eq_thm1_1}-\eqref{eq_thm1_2}, the system is unstable.

%% file: Sections/04_Example.tex
\section{Stability analysis of the network with congestion propagation}
\label{sec_example}

We state the main results as follows:
\begin{thm}
\label{thm_2}
Given Assumptions~\ref{asm_1}-4, the Markov chain \eqref{eq_markovchain_2} with the state space $\mathbb{R}_{\geq0}\times\mathcal{X}_{e_1}\times\mathcal{X}_{e_2}\times\mathcal{D}\times\mathcal{C}$ is stable if there exists a vector $\theta:=[\theta_{e_1},\theta_{e_2}]^{\mathrm{T}}\in[0, 1]^2$ and a positive scalar $\gamma>0$ such that
\begin{align}
    &\alpha- \sum_{e\in\{e_1,e_2\}} (1-\theta_{e})\mathbb{E}_{x_{e_1},x_{e_2}}[q_{e}^{\mathrm{in}}(f_{e_0}(x_{e_0}^c), x_{e_1},x_{e_2}, C)]- \nonumber \\
    & - \sum_{e\in\{e_1,e_2\}}\theta_{e} f_{e}(x_{e}) < -\gamma, ~\forall (x_{e_1},x_{e_2})\in\mathcal{X}_{e_1}\times\mathcal{X}_{e_2}, 
    \label{eq_thm2_1}
\end{align}
where $x_{e_0}^c:=\inf\{x_{e_0}|f_{e_0}(x_{e_0})=Q_{e_0}\}$ and
\begin{align}
    &\mathbb{E}_{x_{e_1},x_{e_2}}[q_{e}^{\mathrm{in}}(f_{e_0}(x_{e_0}), x_{e_1},x_{e_2}, C)] \nonumber \\
    &:= \int_{\mathcal{C}} q_{e}^{\mathrm{in}}(f_{e_0}(x_{e_0}), x_{e_1}, x_{e_2}, c)) \Gamma_{x_{e_1,e_2}}(\mathrm{d}c).
\end{align}
\end{thm}

\begin{thm}
\label{thm_3}
Given Assumptions~\ref{asm_1}-4, the Markov chain \eqref{eq_markovchain_2} with the state space $\mathbb{R}_{\geq0}\times\mathcal{X}_{e_1}\times\mathcal{X}_{e_2}\times\mathcal{D}\times\mathcal{C}$ is unstable if there exists a vector $\theta:=[\theta_{e_1},\theta_{e_2}]^{\mathrm{T}}\in[0,1]^2$ and a non-negative scalar $\gamma\geq0$ such that
\begin{align}
    &\alpha- \sum_{e\in\{e_1,e_2\}} (1-\theta_{e})\mathbb{E}_{x_{e_1},x_{e_2}}[q_{e}^{\mathrm{in}}(f_{e_0}(\bar{x}_{e_0}), x_{e_1},x_{e_2}, C)]- \nonumber \\
    & - \sum_{e\in\{e_1,e_2\}}\theta_{e} f_{e}(x_{e}) \geq \gamma, ~\forall (x_{e_1},x_{e_2})\in\mathcal{X}_{e_1}\times\mathcal{X}_{e_2}, \label{eq_thm3_1}
\end{align}
where $\bar{x}_{e_0}:=\infty$.
\end{thm}

Note that $x_{e_0}^c$ defined in Theorem~\ref{thm_2} is usually interpreted as \emph{critical density} since link $e_0$ with $x_{e_0}>x_{e_0}^c$ is considered as ``congested'' in practice. Theorem~\ref{thm_2} indicates that though link $e_0$ could be congested with extremely high traffic densities, we only need to check the critical density. Besides, Theorem~\ref{thm_3} says that we need to check $x_{e_0}=\infty$, namely to consider $\sup f_{e_0}$. 

One can implement Theorem~\ref{thm_2} by solving the following semi-infinite programming (SIP \cite{stein2012solve}):
\begin{equation}
    (P_1)~ \min_{\theta_1,\theta_2,\gamma}~\gamma~s.t.~\eqref{eq_thm2_1},
\end{equation}
and Theorem~\ref{thm_3} by solving the SIP: 
\begin{equation}
    (P_2)~ \max_{\theta_1,\theta_2,\gamma}~\gamma~s.t.~\eqref{eq_thm3_1}.
\end{equation}
We conclude the Markov chain \eqref{eq_markovchain_2} is stable if the optimal value of $P_1$ is positive and unstable if the optimal value of $P_2$ is non-positive. The programmings $P_1$ and $P_2$ belong to SIPs because they have infinite constraints over the continuous set $\mathcal{X}_{e_1}\times\mathcal{X}_{e_2}$. But noting $\mathcal{X}_{e_1}$ and $\mathcal{X}_{e_2}$ are bounded, we have efficient algorithms to solve $P_1$ and $P_2$ \cite{stein2012solve}.

Note that Theorem~\ref{thm_2} is proved based on the Lyapunov function $\tilde V:\mathbb{R}_{\geq0}^3\to\mathbb{R}_{\geq0}$:
\begin{equation}
    \tilde{V}(x_{e_0}, x_{e_1},x_{e_2}) = x_{e_0}(\frac{1}{2}x_{e_0} + \theta_{e_1}x_{e_1} + \theta_{e_2}x_{e_2}). \label{eq_lyaunov_2}
\end{equation}
and Theorem~\ref{thm_3} is based on the test function $\tilde{W}:\mathbb{R}_{\geq0}^3\to\mathbb{R}_{\geq0}$:
\begin{equation}
    \tilde{W}(x_{e_0},x_{e_1},x_{e_2}) = \xi_1 - \frac{1}{x_{e_0}+ \theta_{e_1}x_{e_1} + \theta_{e_2}x_{e_2} + \xi_2}, \label{eq_lyapunov_3}
\end{equation}
where $\xi_1$ and $\xi_2$ are sufficiently large numbers. Clearly, we can further improve the stability and instability conditions by considering more sophisticated Lyapunov/test functions, such as those replacing the linear term $\theta_{e_1}x_{e_1}+\theta_{e_2}x_{e_2}$ with non-linear terms. However, it incurs more computational costs. 

The following section presents a numerical example. The proofs of Theorems~2 and 3 are omitted since they are similar to those of Theorem~1, except for different Lyapunov/test functions.

\subsection{Numerical example}
\begin{figure}[htbp]
    \centering
    \includegraphics[width=0.6\linewidth]{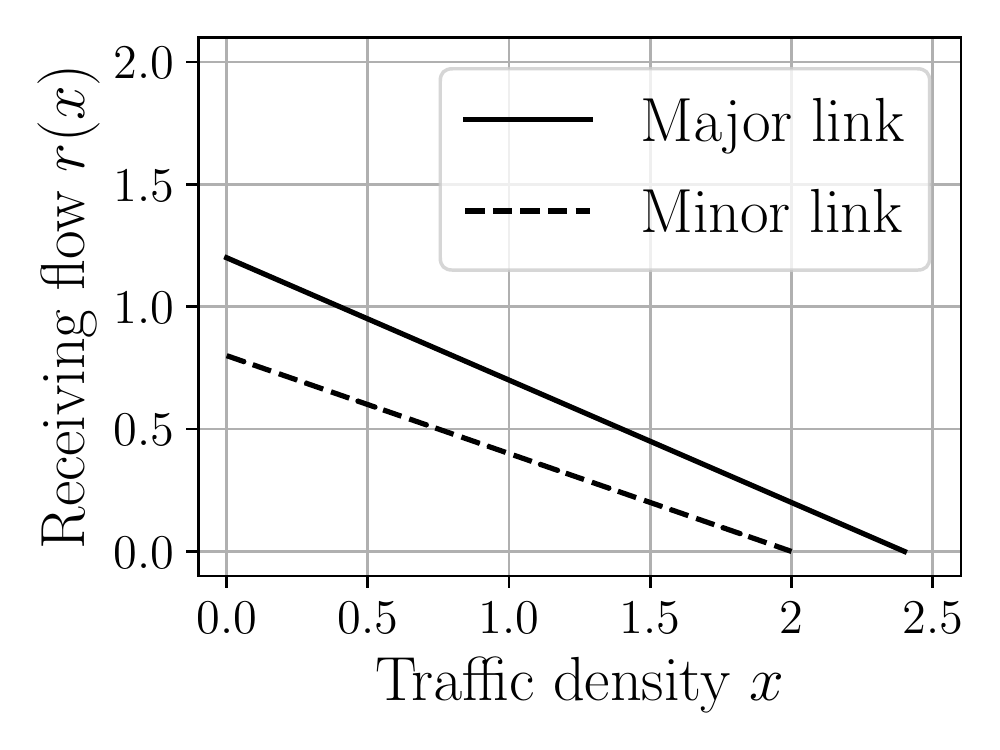}
    \caption{Receiving flows of major and minor links}
    \label{fig_receivingflow}
\end{figure}

Besides the setting in Section~\ref{sec_num_inf}, we introduce the receiving flows
\begin{equation}
    r_e(x_e) = R_e - w_ex_e
\end{equation}
with $R_{e_1}=1.2$, $R_{e_2}=0.8$, $w_{e_1}=0.5$ and $w_{e_2}=0.4$, as illusrated in Fig.~\ref{fig_receivingflow}. It follows $x_{e_1}^{\max}=2.4$ and $x_{e_2}^{\max}=2$. For the upstream link $e_0$, we suppose that its sending flow  with $v_{e_0}=1$ and $Q_{e_0}=1$, which indicates the critical density $x_{e_0}^c=1$.

Analyzing the Markov chain \eqref{eq_markovchain_2} is more difficult since traffic dynamics involving congestion spillback is more complicated. We consider the technique of invariant sets \cite{blanchini1999set} and focus on our analysis of the Markov chain \eqref{eq_markovchain_2} on the state space 
\begin{equation}
    [\underline{d}, \infty)\times [\underline{x}_{e_1}, \bar{x}_{e_1}]\times[0, \bar{x}_{e_2}]\times [\underline{d}, 1.2]\times[0, \bar{c}], \label{eq_invariance}
\end{equation}
where the boundaries $\underline x_{e_1}$, $\bar{x}_{e_1}$ and $\bar{x}_{e_2}$ satisfy 
\begin{subequations}
\begin{align}
    \beta_{e_1}(\underline{x}_{e_1}, 0)\underline{x}_{e_1}\underline{d}+\beta_{e_2}(\underline{x}_{e_1},0)(1-\bar{c})\underline{d} =& f_{e_1}(\underline{x}_{e_1}), \\
    r_{e_1}(\bar{x}_{e_1}) =& f_{e_1}(\bar{x}_{e_1}), \\
    \beta_2(\bar{x}_{e_1}, \bar{x}_{e_2}) f_{e_0}(x_{e_0}^c)\bar{c}=& f_{e_2}(\bar{x}_{e_2}).
\end{align}
\end{subequations}
Note that restricting analysis on the state space \eqref{eq_invariance} does not lose any generality. In fact, we can prove that the set
\begin{equation}
    \tilde{\mathcal{X}}:=[\underline{d}, \infty)\times [\underline{x}_{e_1}, \bar{x}_{e_1}]\times[0, \bar{x}_{e_2}]
\end{equation}
is positively invariant and globally attracting. That is, for any initial condition $(X_{e_0}(0), X_{e_1}(0), X_{e_2}(0))\in\tilde{\mathcal{X}}$, $(X_{e_0}(t), X_{e_1}(t), X_{e_2}(t))\in\tilde{\mathcal{X}}$ for $t\geq0$ given any $(D(t),C(t))\in[\underline{d},1.2]\times[0,\bar{c}]$. Besides, for any initial condition $(X_{e_0}(0), X_{e_1}(0), X_{e_2}(0))\in\mathbb{R}_{\geq0}\times[0,x_{e_1}^{\max}]\times[0, x_{e_2}^{\max}]$, $(X_{e_0}(t), X_{e_1}(t), X_{e_2}$ enters $\tilde{\mathcal{X}}$ almost surely. 

\begin{figure}[htbp]
    \centering
    \begin{subfigure}{0.45\linewidth}
    \centering
    \includegraphics[width=\linewidth]{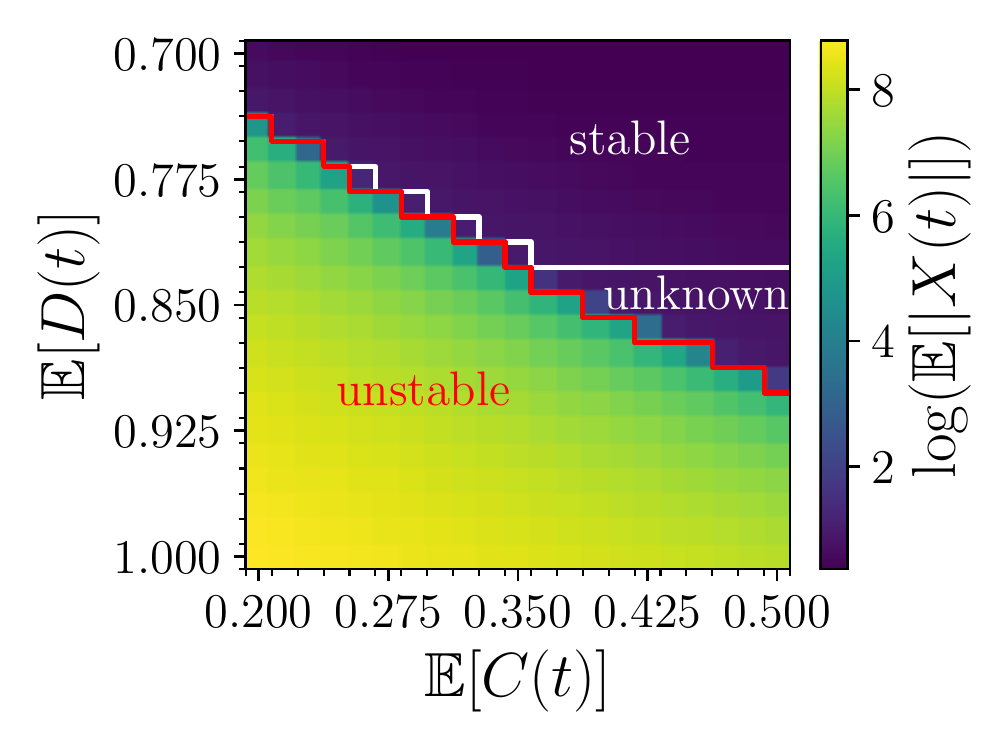}  
  \caption{Stability region.}
  \label{fig_fin_stability}
\end{subfigure}
\begin{subfigure}{0.45\linewidth}
  \centering
  \includegraphics[width=\linewidth]{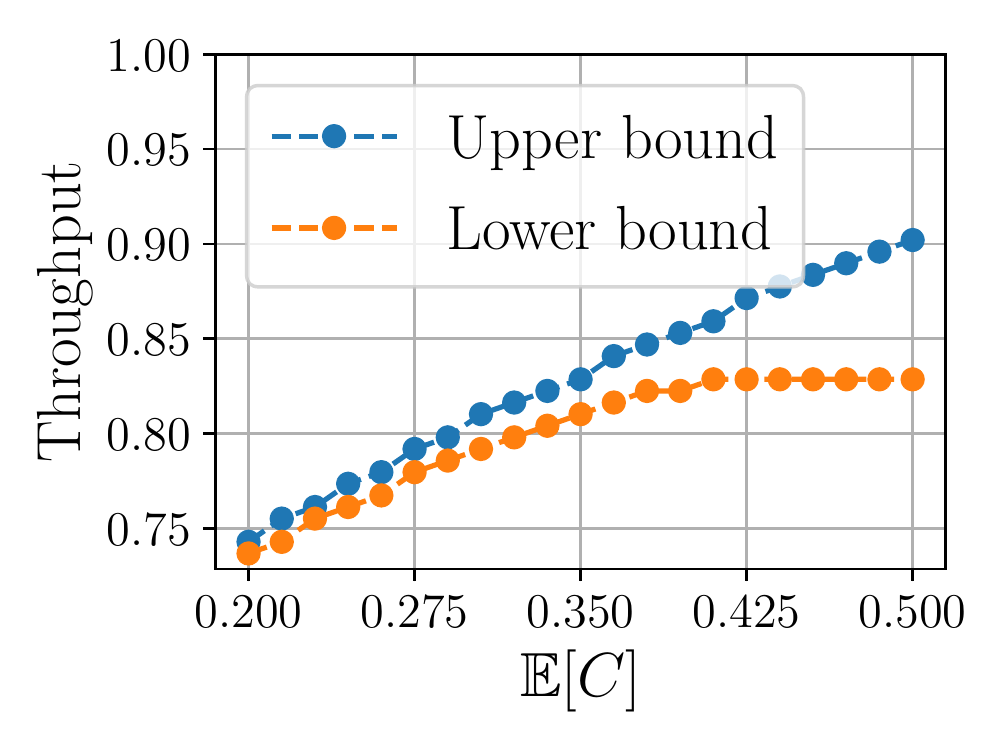}  
  \caption{Throughput.}
  \label{fig_fin_throughput}
\end{subfigure}
    \caption{Analysis of stability and throughput given links $e_1$ and $e_2$ with finite buffer sizes.}
    \label{fig_finite}
\end{figure}

Fig.~\ref{fig_fin_stability} presents the time-average of traffic densities after $5\times10^5$ steps and discloses the stability and instability regions. We have two observations. First, there exists a gap between the stability and instability regions. This is because our stability and instability criteria are only sufficient. Second, the stability region in Fig.~\ref{fig_fin_stability} shrinks significantly, compared with that in Fig.~\ref{fig_region_inf}. It indicates that congestion spillback may not be neglected in analyzing real-world scenarios.

Fig.~\ref{fig_fin_throughput} shows upper and lower bounds of throughput. We note that the gap tends to be enlarged as the compliance rate increases. Per our discussion on Theorems~\ref{thm_2} and \ref{thm_3}, it is possible to narrow down the gap by considering more advanced Lyapunov/test functions.

%% file: Sections/05_Conclusion.tex
\section{Concluding remarks}
\label{sec_conclusion}

In this paper, we considered the traffic stability and throughput of a parallel-link network subject to non-compliant traffic flows. We formulated a Markov chain that captures the traffic evolution under a dynamic routing strategy and in the face of a state-dependent non-compliance rate of drivers. Using Lyapunov methods, we derived stability conditions for several typical settings with or without traffic spillbacks. We also used the results to analytically quantify the impact of driver non-compliance on network throughput.
Possible future directions include extension of the results to general single-origin-single-destination networks with cyclic structures and multi-commodity scenarios.

\appendix
\subsection{Proof of Lemma~\ref{lmm_1}}
\label{app_pf_lmm1}

By Proposition 7.1.4 in  \cite{meyn2012markov}, we see that the system \eqref{eq_inf_1}-\eqref{eq_inf_2} is forward accessible. Then Assumption~4.1 indicates that the Markov chain~\eqref{eq_markovchain_1} is $\varphi$-irreducible by Theorem 7.2.6 in \cite{meyn2012markov}. 

Since we assume the flow functions are continuous (see Assumptions~1-2), there exists a neighborhood such that \eqref{eq_asm4_1}-\eqref{eq_asm4_2} hold. It implies that the system \eqref{eq_fin_1}-\eqref{eq_fin_3} is forward accessible at leat over the neighborhood. Then the Markov chain \eqref{eq_markovchain_2} is also $\varphi$-irreducible.